# HAP-Centric Flying Ad-Hoc Networks with Cell-Free Non-Terrestrial Connectivity

Muhammet Kırık, *Graduate Student Member, IEEE*, Liza Afeef, *Member, IEEE*, Halim Yanikomeroglu, *Fellow, IEEE* and Hüseyin Arslan, *Fellow, IEEE*

***Abstract*—High-altitude platforms (HAPs) are key enablers of next-generation non-terrestrial networks (NTNs), offering wide coverage, long endurance, and rapid deployment. Despite these advantages, current NTN designs remain satellite-centric and rely on terrestrial cellular assumptions, limiting flexibility and scalability. To overcome these limitations, this article proposes a multi-layer HAP-centric flying ad-hoc network (FANET). In this framework, HAPs are integrated with distributed uncrewed aerial vehicles (UAVs) to form a standalone, cell-free (CF) non-terrestrial system capable of autonomous operation. The layered architecture consists of an inter-HAP ad-hoc layer, a HAP-to-UAV cooperative layer, and a UAV-to-ground access layer, collectively enabling aerial connectivity, adaptive coverage, and interference-aware user access. Unique challenges for each layer are analyzed, including inter-HAP connectivity, FANET co-existence with terrestrial networks (TNs), and user access under heterogeneous conditions. Moreover, the article introduces enabling strategies such as fast beam alignment for high data rate connectivity, uncoordinated FANET/TN co-existence, and user localization and environment classification. Validated by three case studies, the discussion also outlines standardization pathways. The results highlight HAP-centric FANETs as a foundation for resilient, scalable, and application-oriented 6G NTN deployments.**

## I. Introduction

The emerging 6th generation (6G) vision emphasizes intelligent, sustainable, and always-on connectivity as its core principle. This vision is formalized in international telecommunication union (ITU)'s International Mobile Telecommunications for 2030 and Beyond (IMT-2030) framework [1], where the 5th generation (5G) usage scenarios are extended to include immersive, massive, and hyper reliable and low-latency communication (HRLLC) services. In addition, new scenarios such as ubiquitous connectivity, artificial intelligence (AI)-enabled communication, and integrated sensing and communication (ISAC) are introduced to realize the comprehensive vision of 6G. Achieving these goals requires advanced technologies and flexible network architectures. Although ongoing research focuses on improving data rates, reliability, spectrum efficiency, and latency, reliable communication in regions lacking terrestrial infrastructure remains a major challenge. In this regard, non-terrestrial networks (NTNs) are viewed as key enablers of global, seamless, and resilient 6G connectivity.

Despite their anticipated role, NTNs have undergone limited progress since their introduction in 3rd generation partnership project (3GPP) Rel-15 [2]. Although concepts such as space-air-ground integrated networks (SAGINs) [3] aim to integrate spaceborne, airborne, and terrestrial segments into a unified multi-layer communication framework, existing NTN designs still largely rely on terrestrial cellular assumptions, which are unsuited to their flexible nature. Moreover, their rigidity has restricted NTNs' scalability and coordination, preventing them to operate as autonomous networks. To overcome these limitations, future NTN architectures must address the unique requirements of next-generation systems. As 6G aims to integrate sensing and communication under a unified framework, a standalone NTN capable of self-sustained operation without terrestrial support becomes vital, particularly in disaster scenarios where it can ensure connectivity, provide backhaul, and enable sensing and situational awareness for adaptive network operations.

A standalone NTN can be established through a combination of spaceborne and airborne platforms, including low earth orbit (LEO) satellites, high altitude platforms (HAPs), and uncrewed aerial vehicles (UAVs). Among these, HAPs operating in the stratosphere at around 20 km altitude offer distinct advantages [4]. Compared to LEO satellites, HAPs enable faster and more cost-effective deployment, reduced propagation loss, and higher signal-to-noise ratio (SNR) for air-to-ground (A2G) communication and sensing. Moreover, their trajectory-free operation allows flexible mission updates and emergency response. Complementing these capabilities, HAPs also possess greater endurance through onboard fuel storage, support for advanced radio frequency (RF) equipment, and improved stability due to larger size and mass in contrast to UAVs.

Leveraging these characteristics, HAPs can operate as aerial base stations (BSs) to maintain connectivity in regions where terrestrial networks (TNs) fail. When deployed collectively, they can form an independent flying ad-hoc network (FANET) capable of supporting a wide range of 6G applications such as disaster recovery, environmental monitoring, large-scale sensing, and global broadband access [5]. However, dense TN deployments, limited shared spectrum, and the progressively shrinking cell structures, commonly envisioned in layered SAGIN architectures, create co-existence and coordination challenges that may hinder these services. To ensure seamless

M. Kırık, and H. Arslan are with the School of Engineering and Natural Sciences, İstanbul Medipol University, Beykoz, 34810 İstanbul, Türkiye; L. Afeef is with the Department of Electronics and Communication Engineering, Yıldız Technical University, Esenler, 34220 İstanbul, Türkiye; H. Yanikomeroglu is with the Department of Systems and Computer Engineering, Carleton University, Ottawa, ON K1S 5B6, Canada. Corresponding Author: Muhammet Kırık, email: muhammet.kirik@medipol.edu.tr.

integration, future NTNs must adopt distributed control and interference management strategies that enable terrestrial and non-terrestrial layers to operate cooperatively. Leveraging their flexibility, HAPs can achieve this through a cell-free (CF) architecture supported by UAVs, providing intelligent, reliable, and globally accessible connectivity for next-generation systems [6], [7].

To address these challenges and opportunities associated with HAPs and to advance the vision of standalone and application-oriented FANETs for 6G, this article provides the following contributions:

- A multi-layer HAP-centric FANET architecture is introduced, where multiple parent HAPs form an ad-hoc aerial backbone and centrally coordinate distributed UAV swarms as access points (APs) under a CF non-terrestrial framework. Unlike conventional layered SAGIN models that often rely on hierarchical coverage extension and inter-platform connectivity, this architecture provides a HAP-centric coordination, removes the conventional cell-centric view, and enables autonomous NTN operation without continuous dependence on terrestrial infrastructure.
- A detailed analysis of the key technical challenges associated with each layer of the proposed FANET is provided to identify critical limitations in inter-HAP connectivity, vertical HAP–UAV coordination, and UAV-to-ground user access.
- Three representative case studies are developed as enabling mechanisms for the proposed architectural vision, encompassing (i) an ISAC-assisted beam alignment design for fast and high data rate transmission across inter-HAP links, (ii) a sparse codebook design for interference mitigation in uncoordinated FANET/TN co-existence scenarios, and (iii) user environment classification enabled by CF-based sensing. Each case study is evaluated under a unified parameterized framework and analyzed through key performance indicators.
- A standardization outlook is provided, and potential pathways are identified for incorporating the proposed HAP-centric FANET framework into future NTN standards.

## II. Cooperative Structure

To understand how HAPs and UAVs can jointly create an intelligent standalone FANET, it is helpful to view the system as a cooperative, layered structure. In this vision, quasi-stationary HAPs located in a certain region provide wide-area coordination and processing capability, while stationary swarms of UAVs located under their corresponding HAPs extend the coverage closer to users with enhanced cell granularity. Together, they form a CF architecture within a FANET, enabling distributed communication, sensing, and control. The framework can be described through three main layers: the HAP-to-HAP ad-hoc layer, the HAP-to-UAV cooperative layer, and the UAV-to-ground access layer, as illustrated in Fig. 1.

### A. HAP-to-HAP Ad-Hoc Layer

At the highest level, multiple HAPs establish direct inter-HAP communication links to form an aerial backbone that connects the entire FANET. This backbone can be viewed as the stratospheric counterpart of the terrestrial fiber network, designed to exchange data and control information among the HAPs at data rates comparable to those achieved through inter-BS fiber-optic connections. Achieving such high capacity over wireless air-to-air (A2A) links requires advanced RF systems capable of operating at extremely high frequency (EHF) bands, including millimeter wave (mmWave) and sub-terahertz (THz) frequencies. Although operating in these bands offers access to large amounts of unused spectrum, it also introduces new challenges, as EHF signals are highly susceptible to attenuation and propagation loss. To counter these effects, MIMO techniques are employed to increase link robustness and spectral efficiency. Among these, beamforming plays a particularly important role in focusing the transmitted energy toward the intended HAP, thereby minimizing loss and interference. Each HAP is therefore envisioned to carry a large uniform planar array (UPA), allowing analog precoding and combining across both azimuth and elevation planes to support the three-dimensional alignment of narrow beams between dynamically positioned aerial nodes.

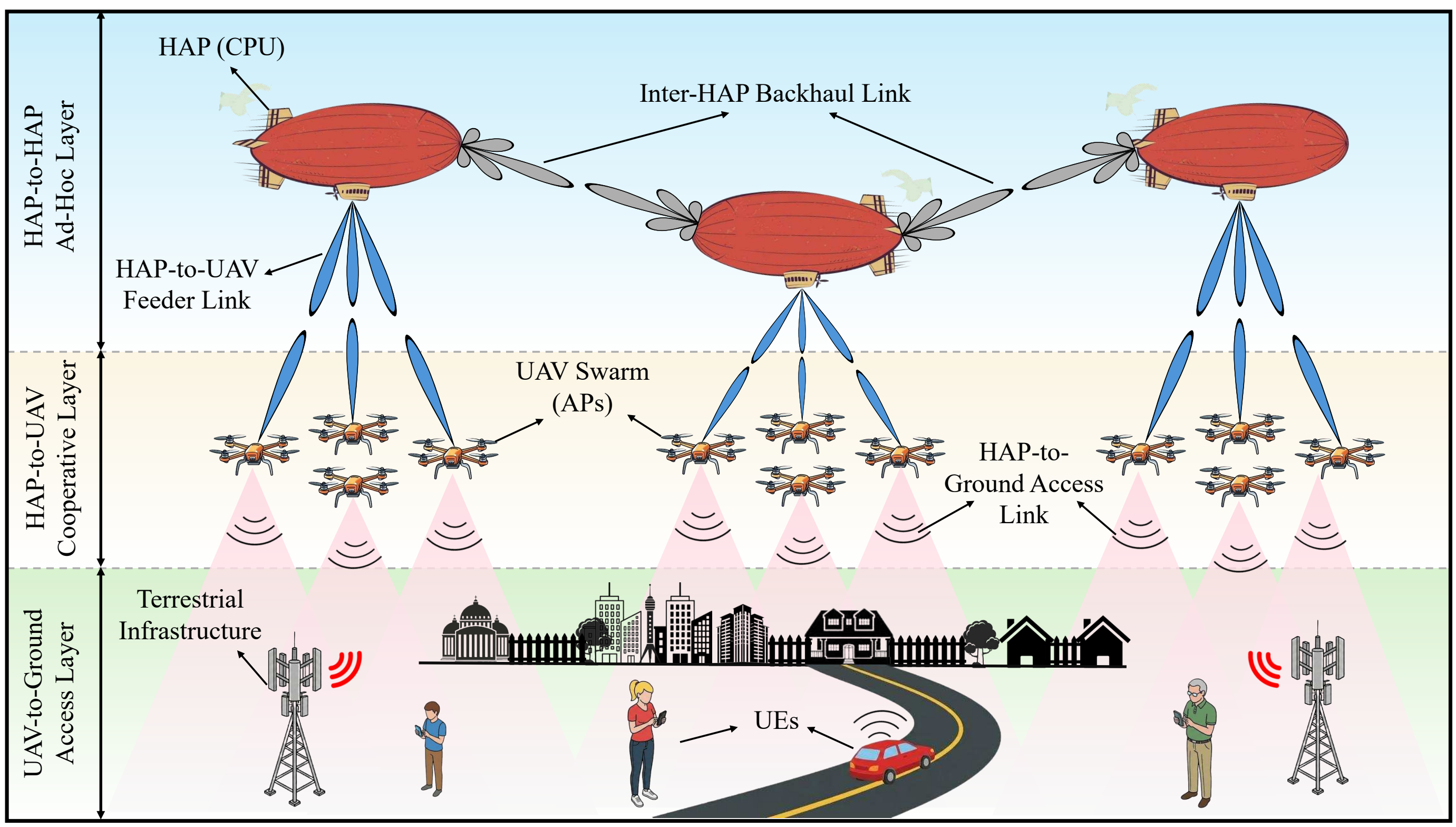

Fig. 1. System architecture of the proposed multi-layer HAP-centric FANET.

### B. HAP-to-UAV Cooperative Layer

The middle layer connects each HAP to a swarm of UAVs flying below it. This layer acts as the vertical bridge that transfers data, sensing information, and control commands between high-altitude and lower-altitude elements. In this relationship, the HAP functions as a central processing unit (CPU), responsible for coordination and decision-making, while each UAV serves as an AP, offering localized access to users. The communication between them also uses directional beams, ensuring high data rates and stable connections even when both nodes are nonstationary. The HAP synchronizes its UAVs, manages their trajectories, and allocates resources among them.

### C. UAV-to-Ground Access Layer

This is the lowest layer representing the interface between the FANET and users on the ground. Operating in complex propagation environments characterized by multipath and shadowing, this layer manages the direct interaction between the aerial network and users. Each UAV provides wireless access for uplink and downlink communication, bringing the network closer to the surface to reduce latency, improve link reliability, and enhance cell granularity. By doing so, a more localized CF connectivity within the FANET is created to reduce the likelihood of overlap with terrestrial cells and enabling seamless co-existence between the two networks. In this configuration, the UAVs handle dense, small-scale user access, while the HAPs manage high-level processing and coordination. Together, these two tiers form a CF architecture to deliver seamless and interference-aware connectivity for ground users.

## III. Design Challenges

While the proposed multi-layered FANET architecture offers a promising foundation for autonomous and resilient NTN operation, its realization introduces several practical and standardization-relevant challenges. These challenges span across the three aerial layers and highlight the need for new 6G mechanisms that extend beyond current NTN capabilities. The key design hurdles are outlined below.

### A. Inter-HAP Connectivity

Inter-HAP connectivity forms the core of the envisioned aerial backbone. However, maintaining high-capacity A2A links across multiple HAPs is inherently challenging due to their quasi-stationary characteristics. Since even small deviations in altitude or movement direction can cause misalignment in highly directional beams, HAPs require frequent re-steering in both azimuth and elevation dimensions. Ensuring stable connectivity therefore demands real-time beam tracking and mobility-aware link control that accounts for each platform's motion and channel dynamics.

Beyond beam alignment, distributed HAP clusters must also maintain tight time–frequency synchronization and support adaptive routing. With no central terrestrial reference, time and frequency offsets can degrade coherent reception, disrupt cooperative processing, and reduce overall spectral efficiency. As the number of HAPs increases, dynamic topology management becomes critical for sustaining backhaul integrity while meeting quality of service (QoS) requirements across a growing airborne network.

These considerations underscore a broader standardization gap: existing NTN specifications do not yet address multi-HAP mesh networking, distributed synchronization, or aerial backhaul routing capabilities that will be essential for scalable stratospheric networks.

### B. FANET/TN Co-Existence

Co-existence with terrestrial networks remains one of the most pressing challenges for low-altitude aerial access. Since A2G links often operate in sub-6 GHz bands to ensure robustness, FANETs and TNs frequently share the same spectrum [8]. This leads to severe co-channel interference (CCI), as UAVs experience line-of-sight exposure to multiple terrestrial cells while simultaneously transmitting toward ground users. Without coordinated spectrum management, this interference can significantly degrade coverage, increase bit error rate (BER), and reduce system capacity.

The challenge becomes even more complex when a user receives identical data streams from both a terrestrial BS and an aerial AP. Due to differing propagation conditions, Doppler characteristics, and timing offsets, the user terminal may struggle to combine these signals effectively. Addressing this issue requires new waveform alignment, synchronization support, and cross-layer link management mechanisms that operate seamlessly across TN and NTN domains.

These co-existence problems highlight the need for NTN-specific interference mitigation frameworks, an area where current 3GPP specifications provide only partial solutions.

### C. User Access

The user-access layer introduces challenges driven by environmental variability and heterogeneous propagation conditions. While UAVs improve link proximity compared to HAPs, users located indoors, in obstructed areas, or in low-SNR environments remain difficult to serve reliably. Deep fading and severe attenuation hinder both communication and sensing performance, reducing the effectiveness of environment classification and degrading user experience.

Delivering robust access in these conditions requires new physical-layer strategies, including adaptive beamforming, energy-efficient waveform design, and sensing-enhanced user localization. Unlike terrestrial BSs, UAVs operate with strict size, weight, and energy constraints, limiting the complexity of onboard hardware and further challenging the design of high-performance access links. These limitations highlight the need for lightweight yet reliable physical-layer solutions and environment aware access strategies as they will play an important role in future NTN standardization efforts.

## IV. Enabling Strategies

The challenges discussed in Section III highlight that realizing a fully autonomous and scalable CF-enabled FANET requires more than isolated technical solutions. Instead, it demands a set of cohesive strategies that jointly support fast link establishment, interference-resilient access, and environment-aware operation across the different layers of the architecture. This section outlines three enabling strategies that address these needs and point toward the capabilities required in future NTN standards.

### A. Fast Beam Alignment for High Data Rate Connectivity

High data rate inter-HAP and HAP–UAV links rely on the ability to maintain narrow-beam connectivity despite mobility and dynamic aerial geometry. Conventional exhaustive or multi-stage search procedures are too slow and resource-intensive for stratospheric mobility. To meet the latency and robustness requirements of the proposed FANET, beam alignment must leverage sensing information to guide the beam search and reduce unnecessary overhead.

A sensing-assisted, dual-stage alignment approach, as in [9], can accelerate the alignment process by combining coarse direction estimation with fine-grained refinement at a higher frequency band. This not only reduces beam training latency but also supports scalable operation as additional HAPs and UAVs join the network. Such approaches highlight the need for enhanced beam management procedures in future NTN standards, particularly those capable of operating autonomously without frequent reliance on terrestrial references.

### B. Uncoordinated FANET/TN Co-Existence

The shared use of sub-6 GHz bands between FANETs and TNs creates unavoidable CCI, especially in scenarios where cross-network coordination is not feasible. Conventional interference mitigation techniques designed for terrestrial cellular systems are insufficient for managing the strong line-of-sight interference patterns seen by aerial APs. As a result, FANETs must adopt transmission schemes that remain robust even in uncoordinated spectrum environments.

A sparse codebook-based transmission approach, as in [10], offers a promising solution by minimizing overlap across spatial and spectral dimensions. By allowing each aerial AP to operate using a sparse signature rather than dense data-based waveforms, interference power at terrestrial receivers can be significantly reduced without requiring explicit coordination. This strategy underscores the need for new NTN-specific coexistence mechanisms that extend beyond traditional terrestrial interference management and support autonomous operation across shared spectrum.

### C. User Localization and Environment Classification

Reliable user access in a FANET requires accurate knowledge of the user's environment, mobility state, and surrounding propagation conditions. Traditional localization techniques depend heavily on uplink feedback and reference signals, which may be unreliable for users located indoors or in heavily obstructed regions. In contrast, the proposed architecture can leverage integrated sensing capabilities and multi-AP diversity to achieve more robust environment awareness.

By exploiting reflected positioning signals at multiple UAVs and fusing measurements at the coordinating HAP as in [11], the network can infer user locations, identify propagation conditions, and classify environments prior to data transmission. This enables adaptive beam selection, interference-aware resource allocation, and more efficient use of airborne access links, all without increasing complexity at the user device. These capabilities align with emerging interest in 6G ISAC and highlight the potential role of multi-layer NTN structures in advancing future positioning and sensing standards.

## V. Case Studies

This section presents three case studies aligned with the enabling strategies discussed in Section IV, illustrating the capabilities of the proposed FANET architecture in addressing key operational challenges. Each study demonstrates a representative scenario highlighting performance gains and design insights for future NTN systems. To ensure consistency and reproducibility, all case studies follow a unified simulation framework aligned with the multi-layer FANET architecture, with key parameters summarized in Table I.

### A. ISAC-Assisted Beam Alignment

Maintaining high-capacity inter-HAP links is essential for forming a reliable aerial backbone. However, due to HAPs' mobility, even small variations in position can disrupt narrow-beam communication at mmWave or sub-THz frequencies. Conventional beam alignment procedures based on exhaustive or iterative search are too slow to effectively track these dynamics. To address this issue, we consider an integrated sensing-assisted beam alignment strategy, as proposed in our prior work [9], illustrated in Fig. 2(a). The method uses low-frequency sensing to obtain coarse estimates of relative direction, delay, and Doppler, which are then used to guide a refined alignment stage at higher frequencies, significantly reducing the search space and computational overhead.

TABLE I
SYSTEM PARAMETERS

| Parameter | | Value |
|---|---|---|
| HAP Altitude | | ~20 km |
| UAV Altitude | | 300 m |
| Number of UAVs | | 8 |
| Number of Transceivers on HAP | | 128 |
| Number of Transceivers on UAVs | | 16 |
| Number of Transceivers on Ground User | | 4 |
| Inter-UAV Spacing | | 50 m |
| Carrier Frequency | Inter-HAP Backhaul Link | 60 GHz |
| | HAP-to-UAV Feeder Link | 10 GHz |
| | HAP-to-Ground Access Link | 2.4 GHz |
| Sensing Frequency | Inter-HAP Backhaul Link | 6 GHz |
| | HAP-to-Ground Access Link | 2.4 GHz |
| Bandwidth | Inter-HAP Backhaul Link | 4 GHz |
| | HAP-to-UAV Feeder Link | 500 MHz |
| | HAP-to-Ground Access Link | 50 MHz |
| Channel Model | Inter-HAP Backhaul Link | One Tap LoS |
| | HAP-to-UAV Feeder Link | One Tap LoS |
| | HAP-to-Ground Access Link | Multi-Tap Rician |
| Waveform | | OFDM |
| FFT Size | | 64 |
| CP Length | | 16 |

Fig. 2(b) evaluates the proposed method in terms of achievable capacity and alignment overhead. Specifically, the capacity versus SNR plot illustrates the impact of angular beam deviation, showing that maximum capacity is achieved under perfect alignment, i.e., $\theta = 0°$, while increasing deviation leads to noticeable degradation. In parallel, the time overhead plot highlights alignment agility, demonstrating that the sensing-assisted approach significantly reduces alignment time compared to conventional exhaustive and iterative search methods. These results confirm that sensing-assisted beam alignment enables faster and more reliable link establishment under mobility, supporting scalable inter-HAP connectivity.

### *B. Sparse Codebook Design for Uncoordinated FANET/TN Co-Existence*

The co-existence of NTNs and TNs in shared spectrum introduces severe CCI, particularly in uncoordinated deployments where no signaling exchange is possible. To address this issue, our previous work in [10] proposes an interference mitigation method, where the system model is illustrated in Fig. 3(a). In this scenario, a terrestrial BS acts as an aggressor interfering with a user connected to an aerial AP in the FANET. The method consists of three stages: (i) subspace-based blind channel estimation of the aggressor, (ii) optimized sparse codebook design leveraging both victim and interference channel state information, and (iii) pilot-assisted transmission with precoding.

Fig. 3(b) evaluates the BER performance of the proposed method. As observed from the figure, conventional orthogonal frequency division multiplexing (OFDM) exhibits an error floor across the considered SNR range due to strong CCI arising from overlapping frequency resource usage between FANET and TN, regardless of the modulation order. In contrast, the sparse codebook approach achieves progressively lower BER with increasing codeword size, $\mathcal{L}_c$, as larger codebooks enable more distinguishable sparse transmission patterns, thereby enhancing detection reliability at the receiver. As a result, the proposed framework mitigates interference by reducing spectral overlap between desired and interfering signals, without requiring inter-network coordination.

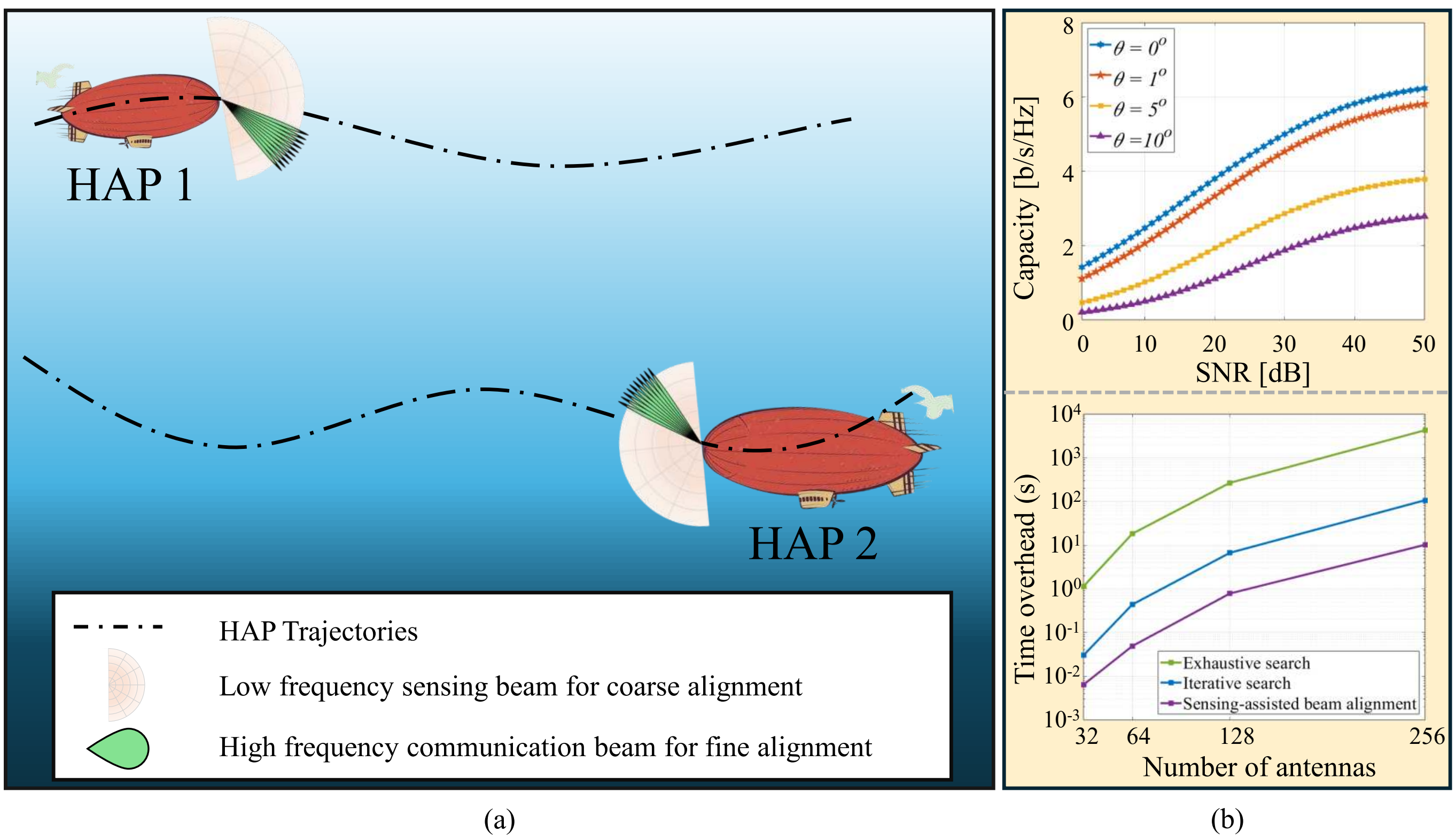

Fig. 2. Case study I: ISAC-assisted multi-layer beam alignment: (a) representative ISAC assisted beam alignment scenario, (b) performance evaluation of the proposed approach in terms of capacity and time overhead.

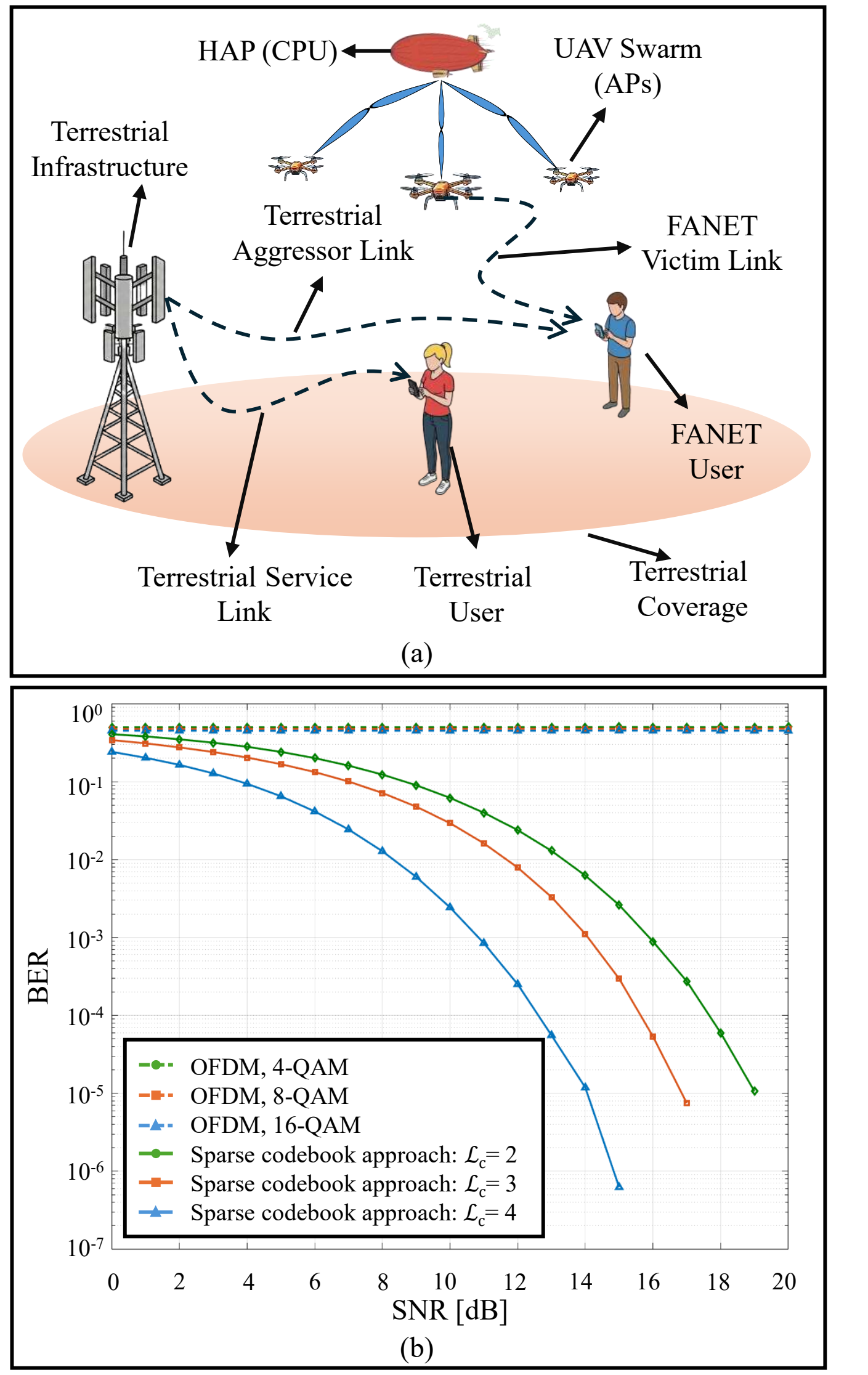

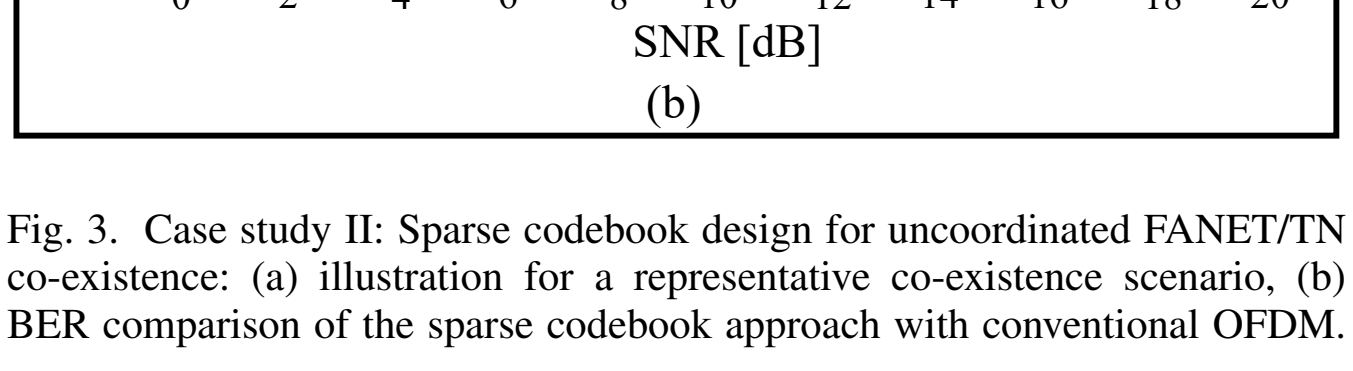

Fig. 3. Case study II: Sparse codebook design for uncoordinated FANET/TN co-existence: (a) illustration for a representative co-existence scenario, (b) BER comparison of the sparse codebook approach with conventional OFDM.

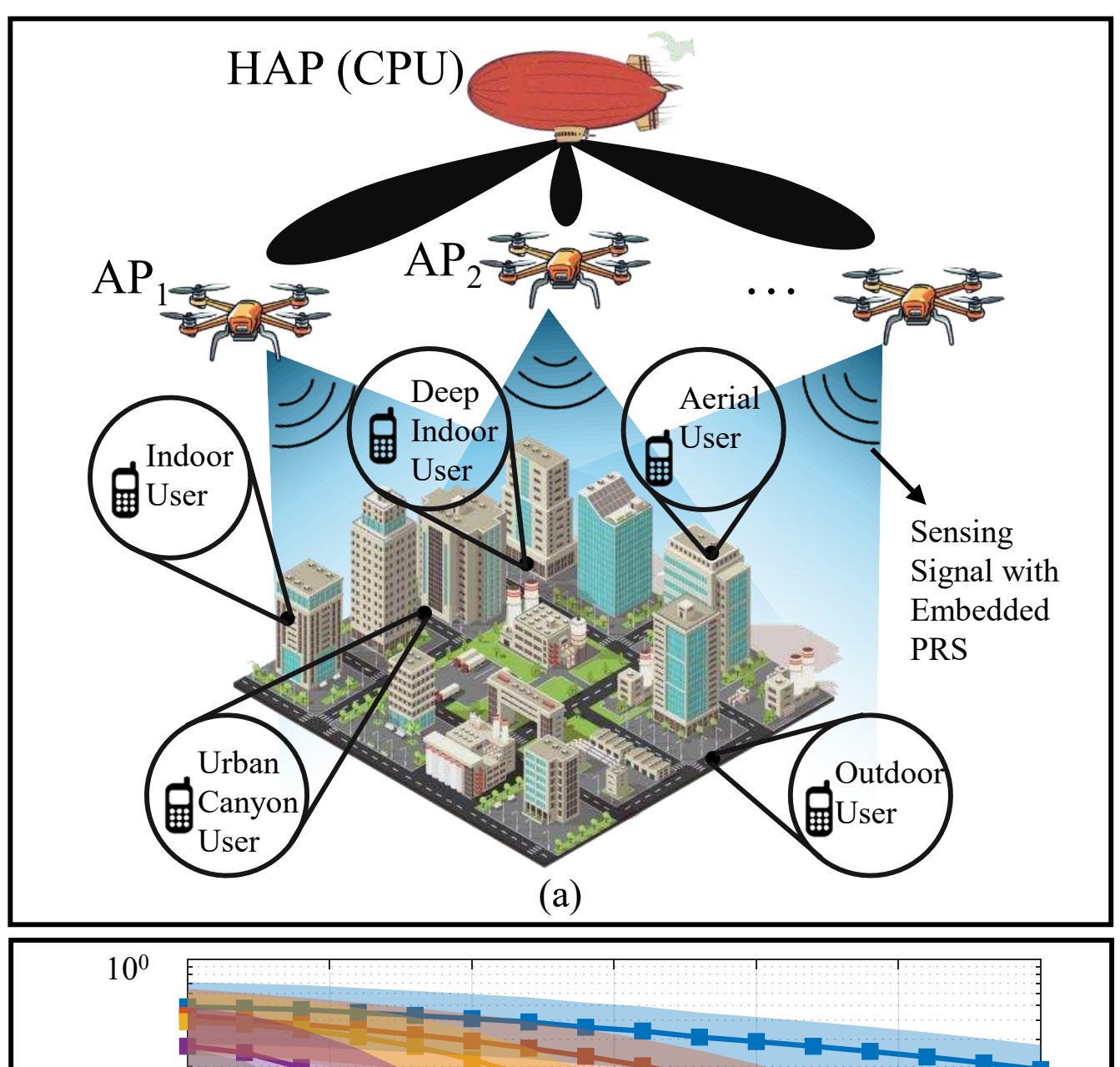

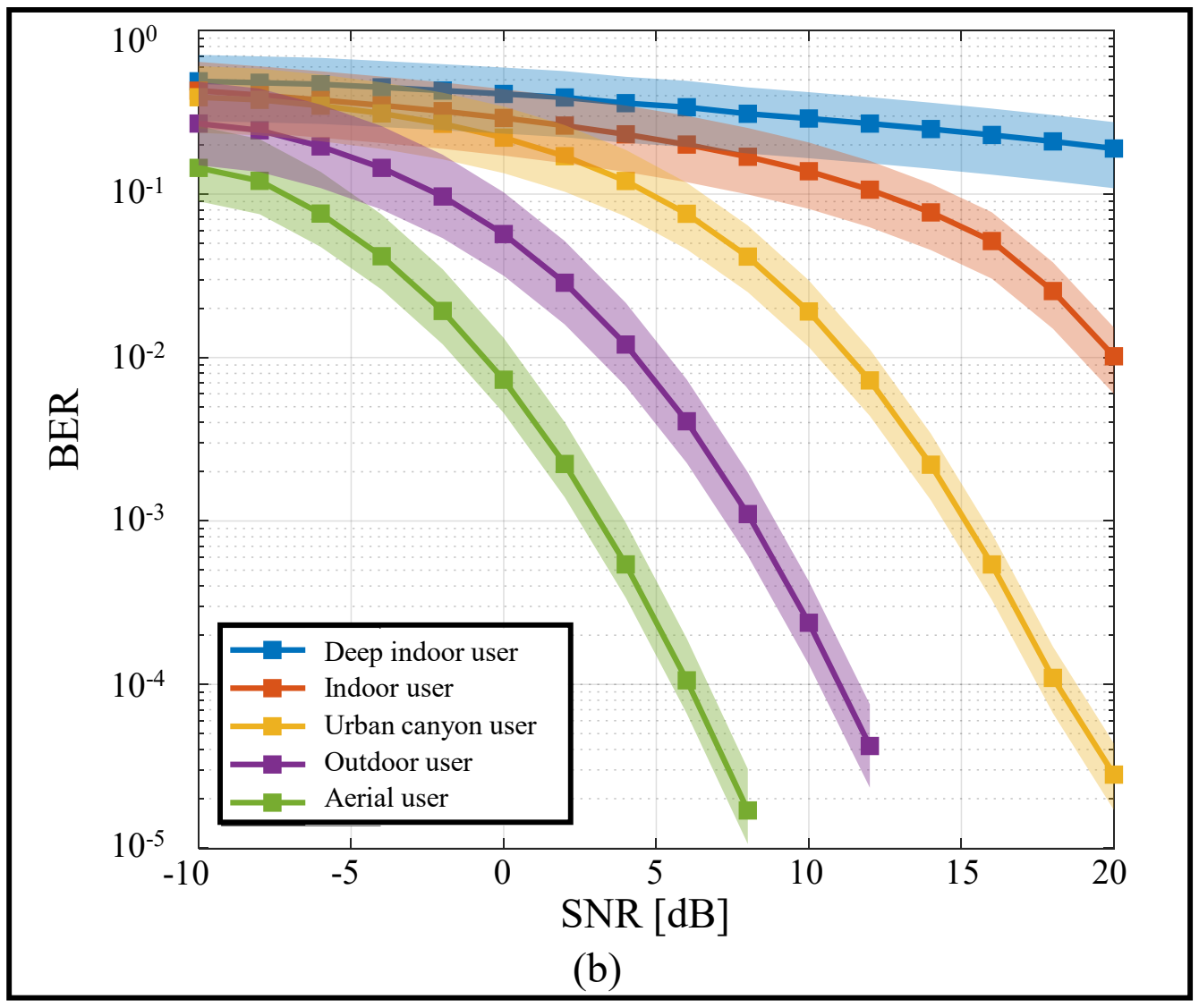

Fig. 4. Case Study III: Sensing-assisted environment classification in CF FANETs: (a) proposed system architecture, (b) BER-based environment classification performance.

### C. CF-Enabled Environment Classification.

A2G links in NTNs exhibit propagation characteristics that vary significantly with the surrounding environment. Accurate classification of the user environment is therefore essential for effective link adaptation and network management. To address this, our previous work in [11] proposes a sensing-assisted classification framework based on a CF architecture, where positioning reference signals (PRSs) are embedded into radar-like transmissions, as illustrated in Fig. 4(a). By exploiting multi-domain diversity, including spatial, frequency, and code domains, across distributed UAVs, the system aggregates the PRS observations across the UAV swarm and applies a BER-based thresholding mechanism to enable reliable environment classification without requiring additional signaling or processing at the user side.

Fig. 4(b) shows the performance of the proposed BER-based classification method across different user environments, including deep indoor, indoor, urban canyon, outdoor, and aerial scenarios, each characterized by distinct propagation conditions. The simulations consider bandwidth aggregation across five propagation frequencies, each employing distinct low-density parity-check (LDPC) codes to enhance sensing diversity. The classification zone thresholds are derived from environment-specific BER statistics based on standardized NTN channel models, in order to maximize separability between adjacent classes. The results demonstrate the effectiveness of the proposed sensing-assisted framework in capturing environment-dependent propagation characteristics, as evidenced by the clear separation between user environments.

## VI. Standardization Pathways

Although NTNs have attracted academic attention for more than two decades, their strategic relevance to industry has increased only recently. This shift has accelerated standardization efforts within 3GPP and ITU, particularly following the introduction of NTN in Rel-17 and its subsequent enhancements in Rel-18 and Rel-19. To ensure that the proposed multi-layer, HAP-centric FANET can evolve from a conceptual design to a deployable architecture, it is essential to examine ongoing standardization activities and highlight potential extensions that could incorporate stratospheric and sub-stratospheric airborne platforms into future NTN and 6G specifications.

### A. *Current Standardization Outlook*

Present NTN standardization focuses primarily on satellite connectivity, with efforts centered on channel models, link budgets, mobility procedures, and waveform adaptations for LEO constellations. While this satellite-oriented foundation is technically robust, it leaves limited room for aerial platforms such as HAPs and UAVs, which operate at vastly different altitudes, mobility regimes, and propagation conditions. As a result, existing specifications do not yet provide explicit guidance for coordinated multi-layer aerial architectures.

Propagation recommendations from the ITU, such as those in [12], offer general models for A2A and A2G links, while [13] introduces baseline NTN architectures for non-terrestrial platforms above 600 km. However, these documents implicitly assume large-footprint cells, long link distances, and satellite-grade Doppler characteristics, which are assumptions that do not extend naturally to stratospheric or low-altitude airborne systems.

Recognizing this gap, recent ITU activities have introduced the HAPs as IMT BSs (HIBS) concept [14] along with stratospheric propagation guidelines tailored for HAPs [15]. These developments represent an important first step toward integrating HAPs into NTN frameworks. Nevertheless, they remain focused on single-layer deployments and do not yet address the cooperative, distributed operation envisioned in multi-layer FANETs. This highlights the need for continued standardization to support more flexible, dynamically coordinated aerial networks.

### B. *Future Directions for Standardization*

Looking ahead to Rel-20 and beyond, standardization must gradually expand beyond satellite-only assumptions to incorporate fully integrated aerial architectures built on HAPs and UAVs. Several technical areas stand out as particularly important for enabling the proposed multi-layer FANET.

The first requirement concerns spectrum allocation. The frequency bands currently identified for NTN, primarily L-band and S-band, are well suited for long-range A2G links but insufficient for the high-throughput inter-HAP and HAP-to-UAV connectivity needed to support narrow-beam, high-capacity backhaul. Future standards should therefore introduce dedicated provisions for stratospheric and sub-stratospheric links, particularly within mmWave and sub-THz bands where large contiguous bandwidths are available.

A second direction involves redefining the concept of cell size and coverage within NTN. Current specifications inherit the large-footprint paradigm from satellite systems, even when applied to HIBSs. Such coverage areas are incompatible with the fine spatial granularity required for interference control and user adaptation in stratospheric deployments. Multi-layer FANETs naturally motivate a transition toward CF operation supported by distributed UAVs, enabling smaller, flexible coverage zones that adapt to user distribution and propagation conditions.

Finally, enhanced co-existence mechanisms with terrestrial systems must be incorporated. Existing co-existence guidelines primarily consider satellite-to-terrestrial interactions and do not reflect the behavior of airborne, low-altitude platforms that may share spectrum with dense terrestrial infrastructure. Future standards should therefore define interference-aware access rules, shared-band operational conditions, and autonomous mitigation procedures tailored to CF aerial NTNs. These extensions will be crucial for ensuring scalable and reliable co-existence between aerial and terrestrial networks as NTN evolve toward more complex multi-layer deployments.

## VII. Conclusion

This article introduced a multi-layer, HAP-centric FANET architecture as a promising foundation for application-oriented NTNs in future 6G systems. By integrating stratospheric HAPs with distributed UAV swarms under a CF framework, the proposed design enables flexible coverage shaping, enhanced user accessibility, and resilient operation independent of terrestrial infrastructure. The architecture was analyzed across three functional layers: inter-HAP ad-hoc connectivity, HAP-to-UAV cooperative control, and UAV-to-ground user access, each revealing distinct design challenges.

To address these challenges, three enabling strategies were presented: sensing-assisted beam alignment for high-capacity aerial links, sparse codebook transmission for autonomous FANET/TN co-existence, and CF-enabled environment classification for improved user access. The effectiveness of these strategies was demonstrated through representative case studies, highlighting their potential to support dynamic, interference-aware, and environment-adaptive NTNs. A review of current standardization efforts and emerging 3GPP/ITU activities further outlined the pathways required for integrating such multi-layer airborne architectures into future wireless standards.

Overall, this article concluded that HAP-centric FANETs emerge as a compelling enabler for resilient, scalable, and intelligent 6G NTN deployments. Future research opportunities include protocol refinement for multi-layer aerial coordination, cross-layer optimization across sensing and communication, and the use of machine learning to support fully autonomous network operation. These directions can accelerate the transition of the proposed vision from conceptual design toward standardized and deployable NTN solutions.

## Biographies

**Muhammet Kırık** [Graduate Student Member, IEEE] (muhammet.kirik@medipol.edu.tr) is currently working toward the Ph.D. degree with Istanbul Medipol University, 34810 Istanbul, Türkiye. His research interests include non-terrestrial networks for 6G and beyond wireless systems.

**Liza Afeef** [Member, IEEE] (lizaafeef92@yahoo.com) is currently an Assistant Professor with the Department of Electronics and Communication Engineering, Yıldız Technical University, 34220 Istanbul, Türkiye. Her research interests include beamforming and MIMO channel modeling for next-generation wireless systems.

**Halim Yanikomeroglu** [Fellow, IEEE] (halim@sce.carleton.ca) is currently a Chancellor's Professor with the Department of Systems and Computer Engineering, Carleton University, Canada. He is also the Director of the Carleton-NTN (Non-Terrestrial Networks) Laboratory. He is a fellow of the Engineering Institute of Canada (EIC), the Canadian Academy of Engineering (CAE), and the Asia-Pacific Artificial Intelligence Association (AAIA). His group's focus is the wireless access architecture for the 2030s and 2040s, and non-terrestrial networks.

**Hüseyin Arslan** [Fellow, IEEE] (huseyinarslan@medipol.edu.tr) is with the Department of Electrical and Electronics Engineering, Istanbul Medipol University, 34810 Istanbul, Türkiye. He has served as a member of the editorial board for IEEE Communications Surveys and Tutorials, IEEE Sensors Journal, IEEE Transactions on Communications, and IEEE Transactions on Cognitive Communications and Networking. He is a Fellow of the National Academy of Inventors and a member of the Turkish Academy of Sciences. His research interests include 6G and beyond radio access technologies.